\documentclass[%
superscriptaddress,
 amsmath,amssymb,
prl,
twocolumn,
]{revtex4-1}
\usepackage[export]{adjustbox}
\usepackage{graphicx}
\usepackage{color}
\usepackage{xcolor}
\usepackage{ulem}
\usepackage{wasysym}
\usepackage{dcolumn}
\usepackage{bm}
\usepackage{hyperref}
\usepackage{url}
\usepackage[mathscr]{euscript}
\usepackage{mathtools}
\definecolor{mygreen}{rgb}{0,0.5,0}
\definecolor{mybrown}{rgb}{0.65,0.16,0.16}
\newcommand{\colg}[1]{\textcolor{mygreen}{#1}}

\newcommand{\colb}[1]{\textcolor{blue}{#1}}
\newcommand{\colr}[1]{\textcolor{red}{#1}}

\newcommand{\colbr}[1]{\textcolor{mybrown}{#1}}

\def\beq {\begin{equation}}
\def\eeq {\end{equation}}
\def\beqa {\begin{eqnarray}}
\def\eeqa {\end{eqnarray}}
\def \bnum {\begin{enumerate}}
\def \enum {\end{enumerate}}
\def\bi {\begin{itemize}}
\def\ei {\end{itemize}}

\def\rel {R_{\lambda}}

\def\la {\langle}
\def\ra {\rangle}

\def\uu{\mbf{u}}

\def\epsm{\la{\epsilon}\ra}

\def\gc{\Gamma(C)}
\def\uu{\mathbf{u}}
\def\uo{\bm{\omega}}
\def\ul{\mathbf{l}}
\def\ua{\mathbf{A}}
\def\pdf{{\mathcal{P}}(\gc)}

\def\gcone{{\Gamma(C_1)}}
\def\gctwo{{\Gamma(C_2)}}
\begin{document}
\title{The area rule for circulation in three-dimensional turbulence}
\author{Kartik P. Iyer}
\affiliation{Department of Physics, Michigan Technological University, Houghton, MI, $49931$, USA}
\affiliation{Department of Mechanical Engineering-Engineering Mechanics, Michigan Technological University, Houghton, MI, $49931$, USA}
\affiliation{Tandon School of Engineering, New York University, New York, NY $11201$, USA}
\author{Sachin S. Bharadwaj}
\affiliation{Tandon School of Engineering, New York University, New York, NY $11201$, USA}
\author{Katepalli R. Sreenivasan}
\email{katepalli.sreenivasan@nyu.edu}
\affiliation{Tandon School of Engineering, New York University, New York, NY $11201$, USA}
\affiliation{Courant Institute of Mathematical Sciences, New York University, New York, NY $10012$, USA}
\affiliation{Department of Physics, New York University, New York, NY $10012$, USA}
\affiliation{Center for Space Science, New York University Abu Dhabi $129188$, United Arab Emirates}

\date{Postprint version of the manuscript published in Proc. Natl. Acad. Sci. U.S.A. {\bf{118}}, e2114679118 (2021)}

\begin{abstract}
An important idea underlying a plausible dynamical theory of circulation in three-dimensional turbulence is the so-called Area Rule, according to which the probability density function (PDF) of the circulation around closed loops depends only on the minimal area of the loop, not its shape. We assess the robustness of the Area Rule, for both planar and non-planar loops, using high-resolution data from Direct Numerical Simulations. 
For planar loops, the circulation moments for rectangular shapes match those for the square with only small differences, these differences being larger when the aspect ratio is further from unity, and when the moment-order increases. The differences do not exceed about $5\%$ for any condition examined here. The aspect-ratio dependence observed for the second-order moment are indistinguishable from results for the Gaussian Random Field (GRF) with the same two-point correlation function (for which the results are order-independent by construction). When normalized by the SD of the PDF, the aspect ratio dependence is even smaller ($< 2\%$) but does not vanish unlike for the GRF. We obtain circulation statistics around minimal area loops in three dimensions and compare them to those of a planar loop circumscribing equivalent areas, and find that circulation statistics match in the two cases only when normalized by an internal variable such as the standard deviation. This work highlights the hitherto unknown connection between minimal surfaces and turbulence. 
\end{abstract}
\maketitle

Universal features of turbulence have been explored most often in terms of the moments of velocity differences over a chosen separation distance, because they play dynamical roles in the interscale energy transfer \cite{MYII} and in the characterization of the transition to turbulence \cite{yakhotdonzis,schuetal,KRS2021}. These so-called structure functions have been shown \cite{Fri95,sreeantonia97} to be multifractal (i.e., moments of each order are governed by independent exponents). Though the progress made has been considerable \cite{KRS2021}, multifractal scaling has turned out to be difficult to understand analytically, and the problem is compounded because high-order structure functions for different velocity components seem to scale differently \cite{RSHT,ISY20}. It is thus reasonable to consider alternatives, one of which is the circulation around closed loops. The circulation around a loop $C$ is defined as
\beq
\label{circ.eq}
\gc = \oint_C \uu \cdot d\ul = \oiint_A \uo \cdot d\ua  \;,
\eeq
where $\uu$ is the velocity, $\bm{\omega}$ (its curl) is the vorticity and $A$ is any area spanning $C$. The second equality is the Stokes' theorem. Migdal \cite{Migdal95,Migdal19a,Migdal19b,Migdal19c,Migdal20}, who provided the first theory of circulation, reasoned qualitatively that the probability density function (PDF) of circulation at large Reynolds numbers should depend uniquely on the minimal area in a universal manner because any simply connected loop has a unique minimal area associated with it. Needless to say, for loops defined on a plane, the minimal area is no different from the classical area.

A recent exploration in \cite{ISY19,Moriconi20,Krstulovic} has indeed shown the attractive simplicity of circulation. A central theme of the circulation studies is the so-called Area Rule, according to which the probability density function of circulation around closed loops depends only on the area of the minimal surface spanned by the loop \citep{umeki,wake,krs96,benzi,ISY19}. This article explores the Area Rule for both planar and non-planar loops, and highlights the role in turbulence dynamics of minimal surfaces, which have been used extensively to model diverse phenomena such as soap films, black holes and protein folding \cite{Terasaki,Chrusciel,Bahmani}.  The article shows the specific ways in which the Area Rule works but also highlights that the magnitude of the minimal area alone is not sufficient to specify the circulation PDF.

We examine the Area Rule using Direct Numerical Simulations (DNS) of the 3D incompressible Navier-Stokes equations. The turbulence is maintained stationary by energy input at the largest scales. Details of the DNS are now standard. The resulting velocity field statistics are given in \cite{ISY17,ISY20} and references therein, and will not be repeated here; only a few details are given in Table~\ref{tab}. The inertial range for the velocity circulation is approximately $r/\eta \in [50,400]$ \cite{ISY17,ISY19}, where $\eta$ is the Kolmogorov scale. $L/\eta$ is roughly the available scale range. We have also constructed a synthetic, divergence-free Gaussian Random Field (GRF) with the same two-point velocity correlation and the available scale range as the DNS data, with the goal of distinguishing kinematic effects from those of the Navier-Stokes dynamics \cite{TVY01}. Table~\ref{tab} also gives some basic parameters of this GRF (for which we use $L/\eta$ only in the sense of an analogue). The vorticity flatness (averaged over the three Cartesian directions) for the DNS is significantly larger than the GRF value of $3$, because of the small-scale intermittency in the DNS field.
\begin{table}
\begin{center}
    \begin{tabular}{|c|c|c|c|c|c|}
    \hline
     & $N^3$ & $L/\eta$ &$R_\lambda$& $u^\prime L$ & $\la \omega^4 \ra/\la \omega^2 \ra^2$ \\
     \hline
    DNS & $8192^3$ &$2276$& $1300$ & $1.8$ & $22.34$ \\
    GRF & $8192^3$ &$2204$& $-$ & $1.7$& $3.00$ \\
    \hline
    \end{tabular}
    \end{center}
    \vspace{2mm}
    \protect\caption{A few characteristics of the DNS and GRF data in a 3D periodic cube with side $L_0=2\pi$ (in length units): $N^3$ is the grid resolution, $L/\eta$ is the available scale range; for the DNS data, $L \approx 0.2 L_0$ is the integral scale and $\eta \equiv (\nu^3/\epsm)^{1/4}$ is the Kolmogorov scale; here, $\nu$ is the kinematic viscosity, $\epsm$ the mean dissipation rate; $\rel$ is the microscale Reynolds number based on the root-mean-square velocity $u^\prime$ and the Taylor microscale $\lambda$, where $\lambda^2 = 15 \nu u^{\prime 2}/\epsm$; $u^{\prime} L$ is the ``large-scale circulation" to be used for various normalization purposes; $\la \omega^4 \ra/\la \omega^2 \ra^2$ is the vorticity flatness for the DNS data and the analogous quantity for the GRF. $\la \cdot \ra$ denotes a volume average over $L_0^3$.
}
\label{tab}
\end{table}
\begin{figure}
\includegraphics [width=0.5\textwidth,center]{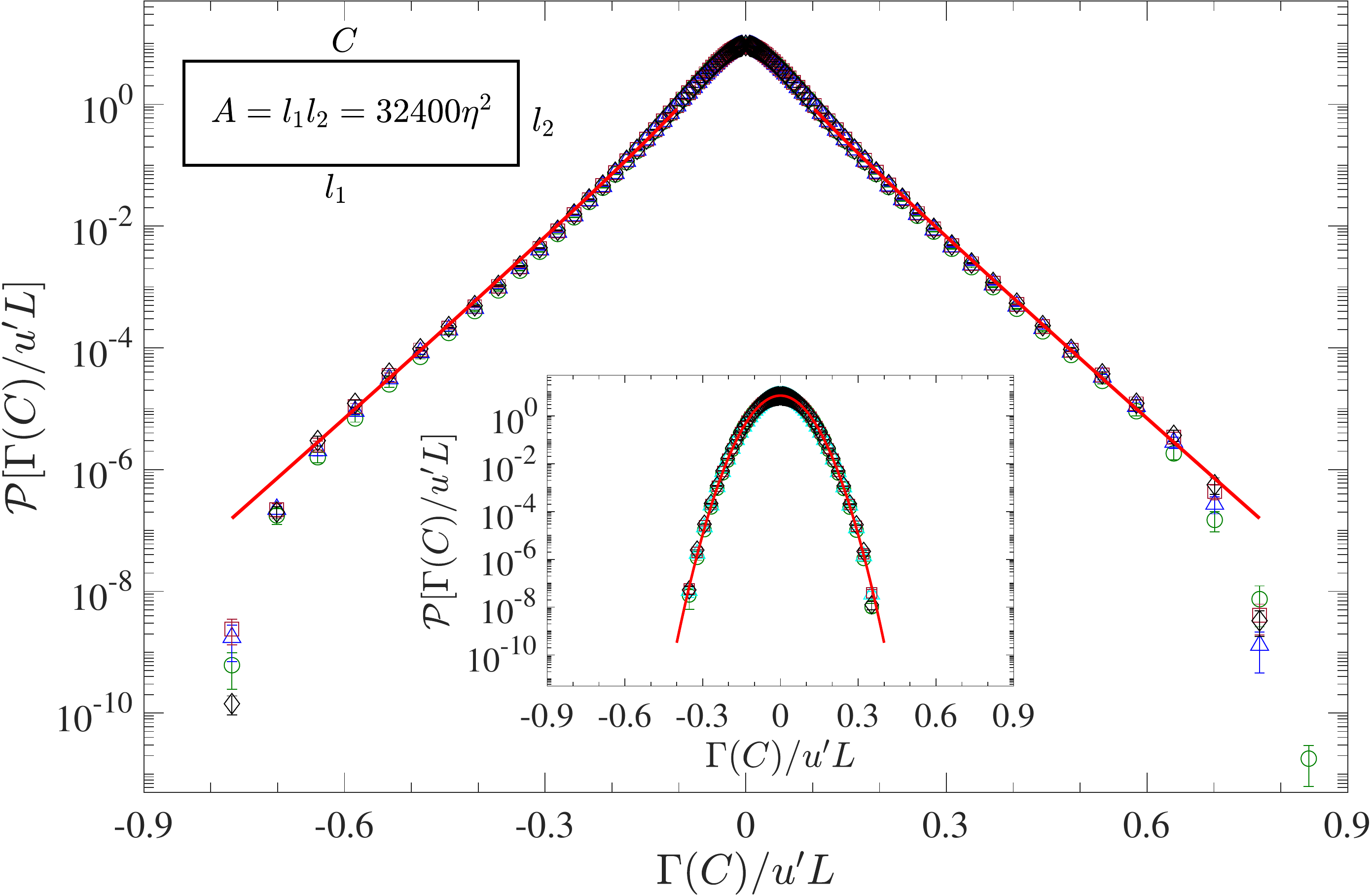}
\protect\caption{
PDF of circulation around a loop $C$ with sides $(l_1,l_2)$ and fixed planar area (top left). Symbols correspond to inertial range dimensions $(l_1/\eta,l_2/\eta)$: $(120,270)$ $(\colg{\circ})$, $(135,240)$ $(\colb{\triangle})$, $(150,216)$ $(\colbr{\Box})$, and $(180,180)$ $(\diamond)$. Error bars indicate $95\%$ confidence intervals from the student-$t$ distribution. Solid line is the PDF fit $\alpha \exp(-b x)/\sqrt{x}$ in $x \in [0.1, 0.7]$ where $x \equiv |\Gamma(C)|/u^\prime L$, $\alpha = 2.49$ and $b = 21.74$. The last two data points corresponding to $< (2\times 10^{-7}) \%$ samples are neglected in the fit. Inset shows corresponding PDFs for GRF with the solid line denoting the Gaussian fit $\alpha_g \exp(-b_g x^2)$ with $\alpha_g = 7.0$ and $b_g =148.7$.
}
\label{pdf.fig}
\end{figure}

We first examine the Area Rule for rectangles in a plane. Figure \ref{pdf.fig} shows the PDFs of circulation around rectangular loops of fixed area A but varying aspect ratios. The PDFs collapse for all rectangles as long as both their sides are contained within the inertial range. The two data points with the lowest probability in Fig.~\ref{pdf.fig} correspond to $1000$ samples or fewer, corresponding to $< 2\times 10^{-7} \%$ of the total number. Neglecting those points, it seems that the inertial range collapse for large $|\Gamma(C)|$ can be fitted quite well by $\alpha \exp(-b|\gc|)/\sqrt{|\gc|}$ \cite{Migdal20}, as shown in Fig.~\ref{pdf.fig} (this fit does not change even when, say, the extreme three data points are also neglected). 

Actually, the circulations PDFs are slightly skewed; e.g., the skewness factor is $0.02$ and the hyperskewness is about $0.5$. We had therefore fitted earlier \cite{ISY19} the two sides of the PDF by stretched exponentials with slightly different stretch factors. But the asymmetry is caused mostly from the core of the distribution; for example, the odd moments decrease with increasing order: $\la \Gamma(C)^m \ra/(u^{\prime}L)^m$ for $m = 3, 5, 7$ and $9$ are $2.6 \times 10^{-6}, 2.3 \times 10^{-7}, 2.8 \times 10^{-8}$ and $4 \times 10^{-9}$, respectively. Thus, we believe that the fit to the tails of the distribution are effectively as stated above, with no need to consider the two sides separately. We note in passing that the origin of the non-zero skewness of circulation, speculated \cite{Migdal95,Migdal19a,Migdal19b,Migdal19c} to be linked to the skewness of longitudinal velocity increments, remains to be better understood.

The PDF of GRF, being Gaussian by definition, possesses very few high amplitude events, as the inset shows. The scaling of all its moments are controlled by that of the second order, unlike turbulence. For this reason, it serves certain comparison purposes well, as discussed below. 
\begin{figure}
\begin{center}
\includegraphics [width=0.5\textwidth,center]{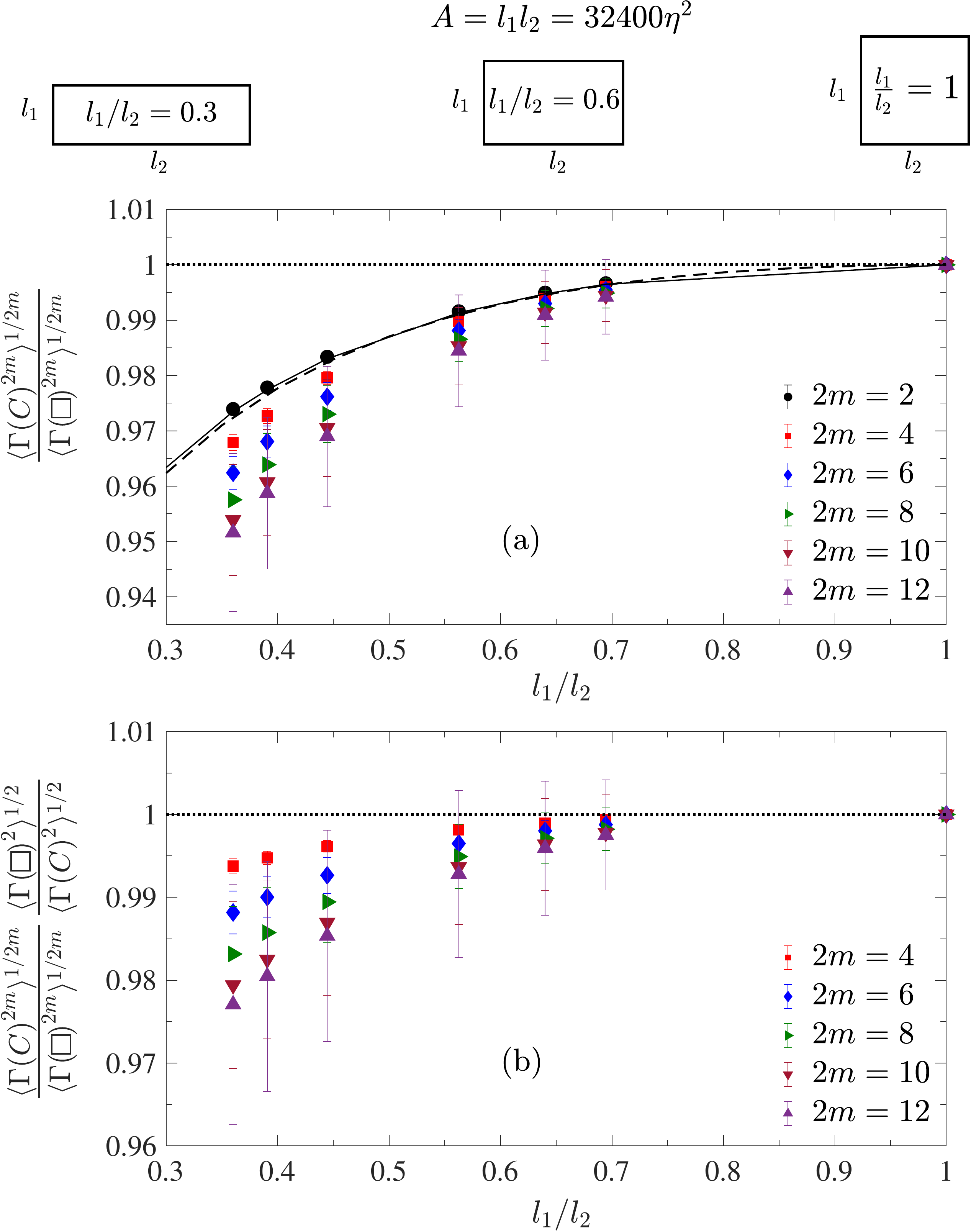}
\end{center}
\protect\caption{
(a) Ratio $Q_{2m}$ of the circulation moment $\la {\Gamma(C)}^{2m} \ra^{1/{2m}}$ around loop $C$ at different aspect ratios (shown in top panel) to that around a square loop $\la {\Gamma(\Box)}^{2m} \ra^{1/{2m}}$ with same area $A$, plotted against the aspect ratio $l_1/l_2$ on linear scales, for orders $2m=2,4,6,8,10$ and $12$. The dimensions of all rectangles considered here fall within the inertial range but with decreasing aspect ratio $l_1$ and $l_2$ approach the dissipative and the large-scale ranges, respectively. The dotted line at unity corresponds to Area Rule holding perfectly. Dashed line is the estimate for $Q_2$ (see Eq.~\ref{gamma2.eq}) while the continuous line is the GRF result, which is order-independent by construction. (b) Ratio $Q_{2m}$ normalized by its standard deviation $Q_2$ plotted against aspect ratio $l_1/l_2$ for $2m=4,6,8,10$ and $12$. The GRF data for all orders corresponds to the dotted line at unity. 
}
\label{momAR.fig}
\end{figure}

While the PDFs in Fig.~\ref{pdf.fig} appear to collapse, it is hard to rule out the existence of small systematic differences because of how strongly the probability axis has been compressed. If there is strict collapse in Fig.~\ref{pdf.fig}, the circulation moments should be independent of the loop aspect ratio for a fixed area.
We examine this by considering in Fig.~\ref{momAR.fig} rectangular loops with different aspect ratios but same fixed area $A$ with $A/\eta^2 \gg 1 $, so that the the loop dimensions $l_1,l_2$ are inside the inertial range.
For $l_1/l_2 =1 $ we have a square loop, but as the aspect ratio $l_1/l_2$ decreases with
$A$ fixed, $l_1$ decreases and approaches the ultraviolet end of the inertial range and $l_2$ increases and approaches the infrared end. 
Figure \ref{momAR.fig}(a) shows the ratio $Q_{2m}$ of the circulation moment $\la {\Gamma(C)}^{2m} \ra^{1/{2m}}$ for a loop $C$ of area $A$ to that for a square loop with the same area $A$, plotted against the loop aspect ratio for $(l_1,l_2)$ within the inertial range. If the Area Rule is exact, $Q_{2m}$ should fall on a horizontal (dotted) line at unity for all orders $2m$. It is clear that they do not: the differences depend on aspect ratio and the moment order, though they are no more than $5\%$ for the cases considered.  


One can also obtain $Q_2$ from the scaling relation for the second-order velocity differences, and thus gain some insight into the aspect ratio dependence. Starting with the relation \cite{Migdal19a,Migdal19b,Migdal19c},
\beq
\label{gamma2.eq}
\la \gc^2 \ra = \oint_C dr_i \oint_C dr^\prime_j \la u_i(r) u_j(r^\prime)\ra \;,
\eeq
we substitute for inertial separations $\eta \ll |r-r^\prime| \ll L$, $ \la u_i(r) u_j(r^\prime)\ra - \delta_{ij} {u^\prime}^2 \propto \delta_{ij}|r-r^\prime|^{\zeta_2}$, where $\zeta_2$ is the scaling exponent for the second-order velocity structure function. Thus while the second-order inertial range circulation exponent is solely determined by $\zeta_2$, the prefactor depends on the shape of the loop, yielding an aspect ratio dependence of $\la \gc^2 \ra$. Indeed, this second-order inertial-range estimate, calculated using $\zeta_2 = 0.72$ and experimental prefactors from \cite{ISY20} changes with changing aspect ratio and compares well with the direct calculation using Eq.~\ref{circ.eq}. This result, shown by the dashed line in Fig.~\ref{momAR.fig}(a), suggests that the moments depend on the shape of the loop $C$ only through the prefactors.  Figure~\ref{momAR.fig}(a) shows that this dependence is small (of the order of $5\%$ across the inertial range). Unfortunately, one cannot 
exploit the tensor forms of higher order velocity correlations to similarly obtain higher order circulation moments.

This modest aspect ratio dependence for the second-order is almost identical to that of circulation around loops in GRF, as Fig.~\ref{momAR.fig}(a) shows (full line). The GRF line shows systematic departure from unity as $l_1/l_2$ decreases, identical to the DNS data using Eq.~\ref{circ.eq}. Note that the GRF result is independent of the moment-order.

The natural question is how much of the departure seen for high-order moments is related to dynamics and how much simply to the GRF behavior. To examine this, we plot the moments $Q_{2m}$ normalized by the standard deviation $Q_2$, which compares the normalized moment $\la {\Gamma(C)}^{2m} \ra^{1/{2m}}/\la {\Gamma(C)}^{2} \ra^{1/{2}}$ of loop $C$ with $\la {\Gamma(\Box)}^{2m} \ra^{1/{2m}}/\la {\Gamma(\Box)}^{2} \ra^{1/{2}}$ for a square loop, both with the same area $A$. 
Figure \ref{momAR.fig}(b) shows this normalized circulation moment $Q_{2m}/Q_2$ as a function of aspect ratio for different orders. The GRF data here fall on the dotted line at unity and are independent of the loop aspect ratio. The normalized circulation moments $Q_{2m}/Q_2$ show a stronger tendency to saturate at unity at all orders examined for aspect ratios $l_1/l_2 > 0.5$ which correspond to
loops with both dimensions well within the inertial range. As the aspect ratio $l_1/l_2$ decreases with $l_1$ and $l_2$ approaching the ends of the inertial range, finite size effects result in the slight deviation from unity, of the order of $2\%$, say, for $Q_{12}/Q_2$. 

\begin{figure}
\includegraphics [width=0.5\textwidth,center]{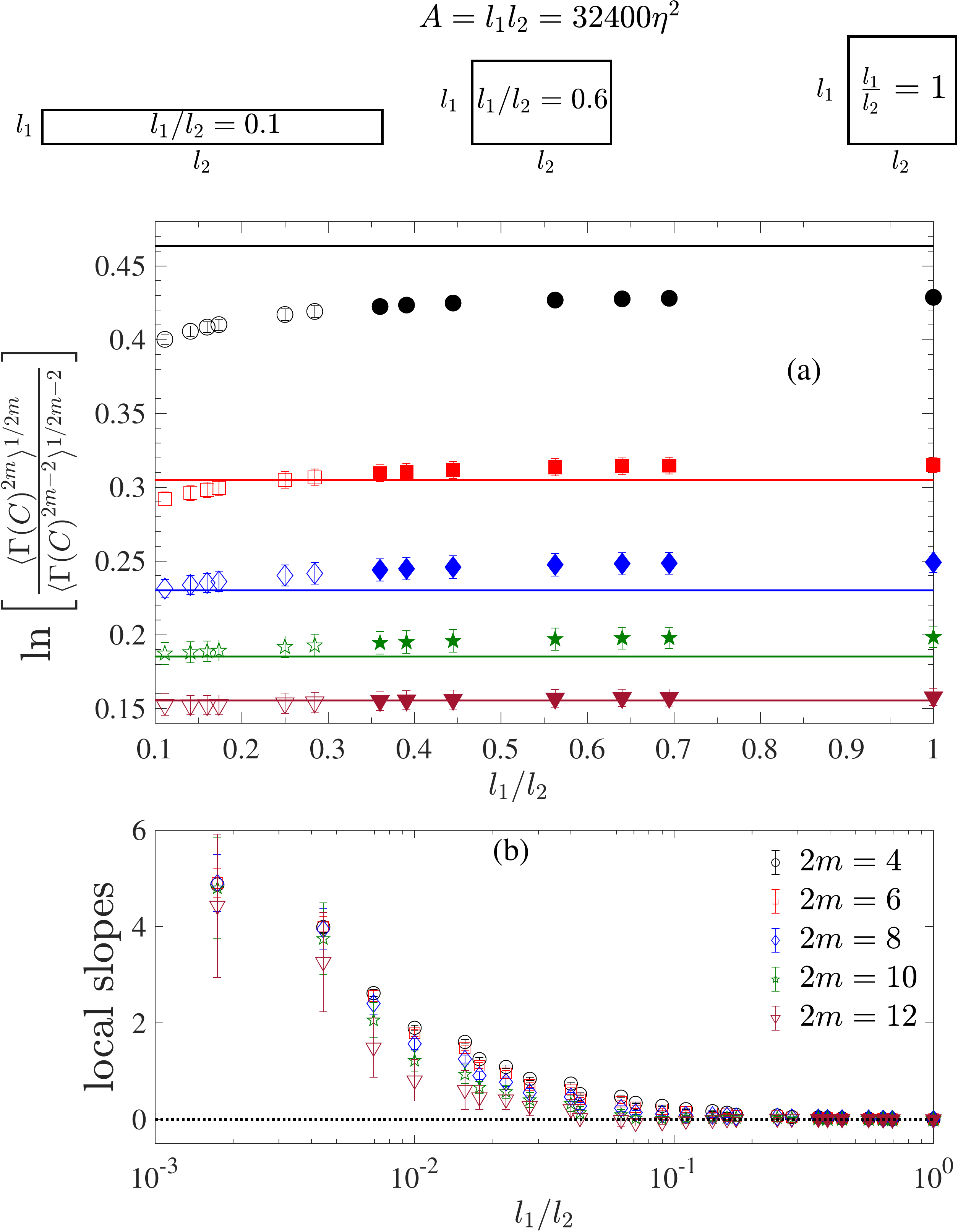}
\protect\caption{
(a) Logarithm of the ratio of normalized circulation moments for loop $C$ plotted against aspect ratio $l_1/l_2$ for loops with sides $l_1$, $l_2$ with fixed area $A$ but varying aspect ratios (see top panel).
Filled symbols correspond to different moment orders $2m = 4 (\bullet),6 ({\colr\blacksquare}), 8 (\colb{\blacklozenge}),10 (\colg{\bigstar})$ and $12 (\colbr{\blacktriangledown})$ for $(l_1,l_2)$ inside the inertial range, while open symbols correspond to rectangles with at least one side outside the inertial range. Horizontal lines corresponds to the high-order estimate from  Eq.~\ref{stirling.eq} which agrees with the DNS data at the largest order $2m=12$ shown here.  
(b) Local slopes from (a) plotted against aspect ratio $l_1/l_2$ on
log-linear scales to show the approach of the higher order moment ratios towards the GRF value of zero which is shown by the horizontal dotted line. 
}
\label{ratiouniv.fig}
\end{figure}

An additional observation from Figs.~\ref{momAR.fig}(a),(b) is that, with increasing order $2m$, $Q_{2m}$ decreases from unity but, for any chosen aspect ratio, this decreasing tendency appears to slow down for $2m \ge 8$. The possibility that $Q_{2m}$ saturates for any given $l_1/l_2$ for large enough order suggests that some scale invariance will be reached by the PDF tails. To explore this further, we plot the ratio of successive even-order moments for a given loop $C$ as a function of the loop aspect ratio, i.e.~for various rectangles $C$ circumscribing the same fixed area as shown in top panel in Fig.~\ref{ratiouniv.fig}. 
The logarithm of this ratio is plotted in Fig.~\ref{ratiouniv.fig}(a) against the aspect ratio $l_1/l_2$ for different loops. In order to study the order-dependence of the circulation moment ratios we consider aspect ratios $l_1/l_2$
in the range $[0.1,1]$. The smaller aspect ratios such as $l_1/l_2=0.1$ correspond to slender rectangles with the loop dimensions lying outside the inertial range. The lower-order ratios show a clear aspect ratio dependence as a rectangle enters the inertial range (loops within inertial range are shown by filled symbols in the figure) while the higher order ratios become independent of aspect ratio even for loops outside the inertial range. In contrast, without showing the result explicitly, we state that GRF shows aspect ratio independence (on this plot) for all orders (approximately at values of $0.27$, $0.18$, $0.13$, $0.10$ and $0.085 $ for $2m = 4,6, 8, 10$ and $12$, respectively). It is noteworthy that the aspect ratio invariance for higher orders appears to extend to loops with edge lengths well outside the inertial range. 

Figure \ref{ratiouniv.fig}(b) plots the derivatives or the local slopes of the logarithm of the successive circulation moment ratios shown in Fig.~\ref{ratiouniv.fig}(a) for different orders as a function of $l_1/l_2$. If the moment ratios shown in \ref{ratiouniv.fig}(a) are indeed constant, such as those for the GRF, with changing aspect ratio, then the corresponding local slopes should be zero for any aspect ratio. It is clear that the circulation moment ratios for the DNS approach aspect ratio invariance faster for higher orders compared to lower orders, with the local slopes reaching zero, for aspect ratios $l_1/l_2$ around $0.1$, at order $12$. 

Since higher order moments largely emerge from the tails of the PDF $\pdf$, which can be fitted quite well by the modified exponential (see Fig.~\ref{pdf.fig}), $\la {\Gamma(C)}^{2m} \ra$ for large even-order $2m$ can be approximated (assuming a symmetric distribution) by
\beq
\label{largem.eq}
\la {\Gamma(C)}^{2m} \ra \approx 2\alpha \int_0^\infty \Gamma^{2m-\frac{1}{2}} e^{-b\Gamma} d\Gamma = 
\frac{2\alpha}{b^{2m+\frac{1}{2}}}{G}(2m+\frac{1}{2}),
\eeq
where $G(z)$ is the Gamma function of $z$. For large even orders $2m$ and $2n$, the Stirling approximation gives
\beq
\label{stirling.eq}
\ln\frac{\la {\Gamma(C)}^{2m} \ra^{1/2m}}
{\la {\Gamma(C)}^{2n} \ra^{1/2n}}
\approx
\ln \Big(\frac{m}{n}\Big) +\frac{1}{4m}\ln(4\pi m) - \frac{1}{4n}\ln(4\pi n). 
\eeq
We note that this estimate is independent of the loop area and the prefactors $\alpha$, $b$ in the PDF fits but depends only on the moment orders $m$ and $n$ (for large enough $m$ and $n$)---and hence, in this sense, universal. The high order approximation shown for $2m=12, 2n = 10$ in Fig.~\ref{ratiouniv.fig}(a) agrees well with the DNS data and confirms the universal nature of the higher order normalized circulation moments. In short, the Area Rule holds for the tails of the normalized circulation distribution as stated in Migdal's papers \cite{Migdal95,Migdal19a,Migdal19b,Migdal19c}.
\begin{figure}
\includegraphics [width=0.5\textwidth,center]{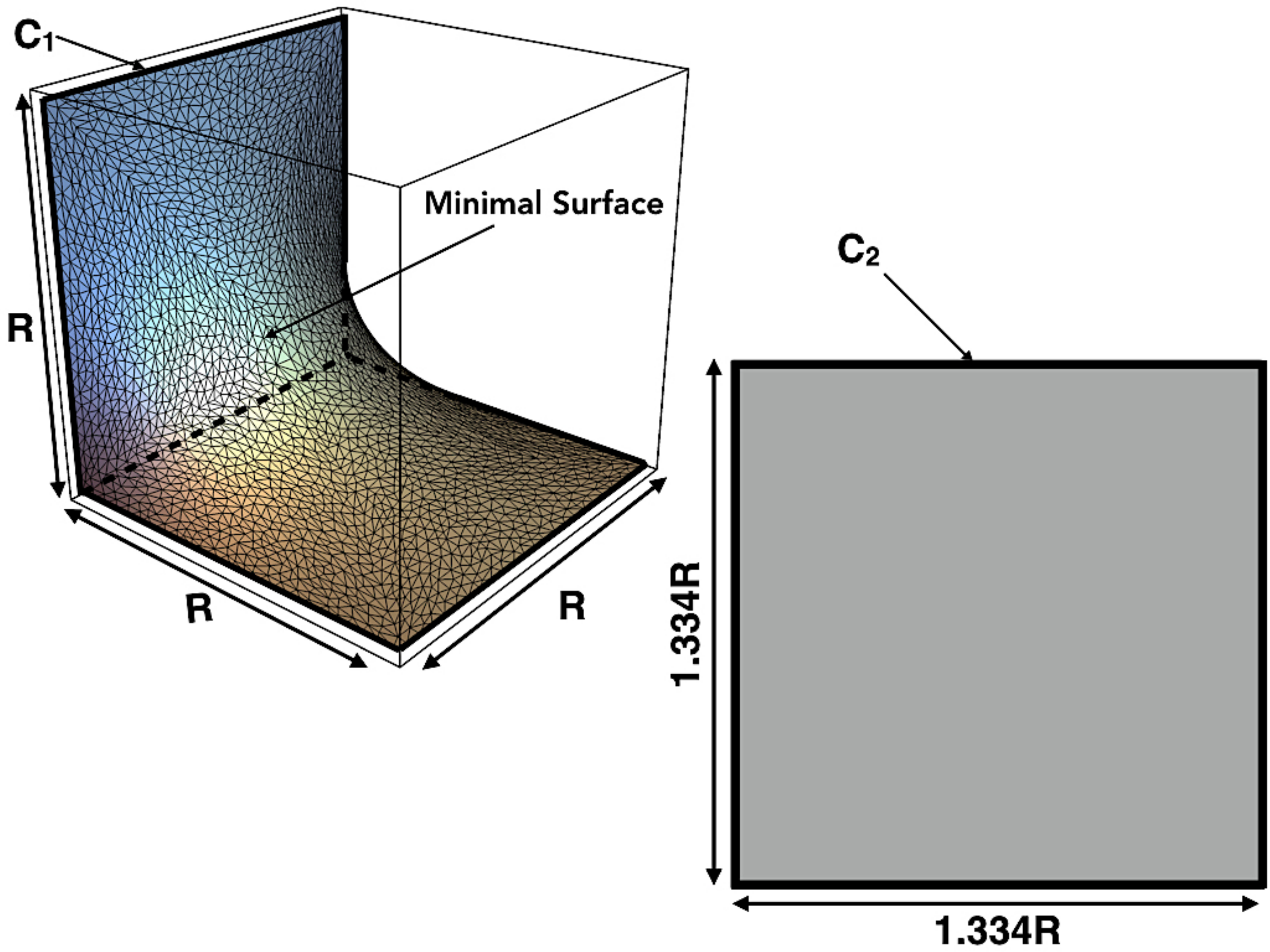}
\protect\caption{
Contour $C_1$ is a non-planar contour with side $R$ and minimal area $A_{C_1} = 1.78 R^2$. Contour $C_2$ is a square with the same planar area $A_{C_2} = 1.78 R^2$. 
}
\label{sg.fig}
\end{figure}

We now examine the Area Rule for non-planar loops. Figure \ref{sg.fig} shows a schematic of a non-planar loop $C_1$ which is a combination of two orthogonal squares, each of side $R$. The minimal surface bounded by $C_1$, having zero mean curvature, is also shown. Starting from an initial guess of a surface that is discretized using triangular meshes, the gradient descent method \cite{pinkall,schumacher} is used to minimize the mean curvature and compute the triangulated minimal surface, using Mathematica\textsuperscript{\textregistered}. Its area is approximately $1.78 R^2$. If the hypothesis is right that the minimal area is what matters for the Area Rule (planar surfaces are trivially minimal), we should be able to compare $\Gamma(C_1)$ with $\Gamma(C_2)$, where $C_2$ is a planar loop of area $1.78 R^2$. 
\begin{figure}
\includegraphics [width=0.5\textwidth,center]{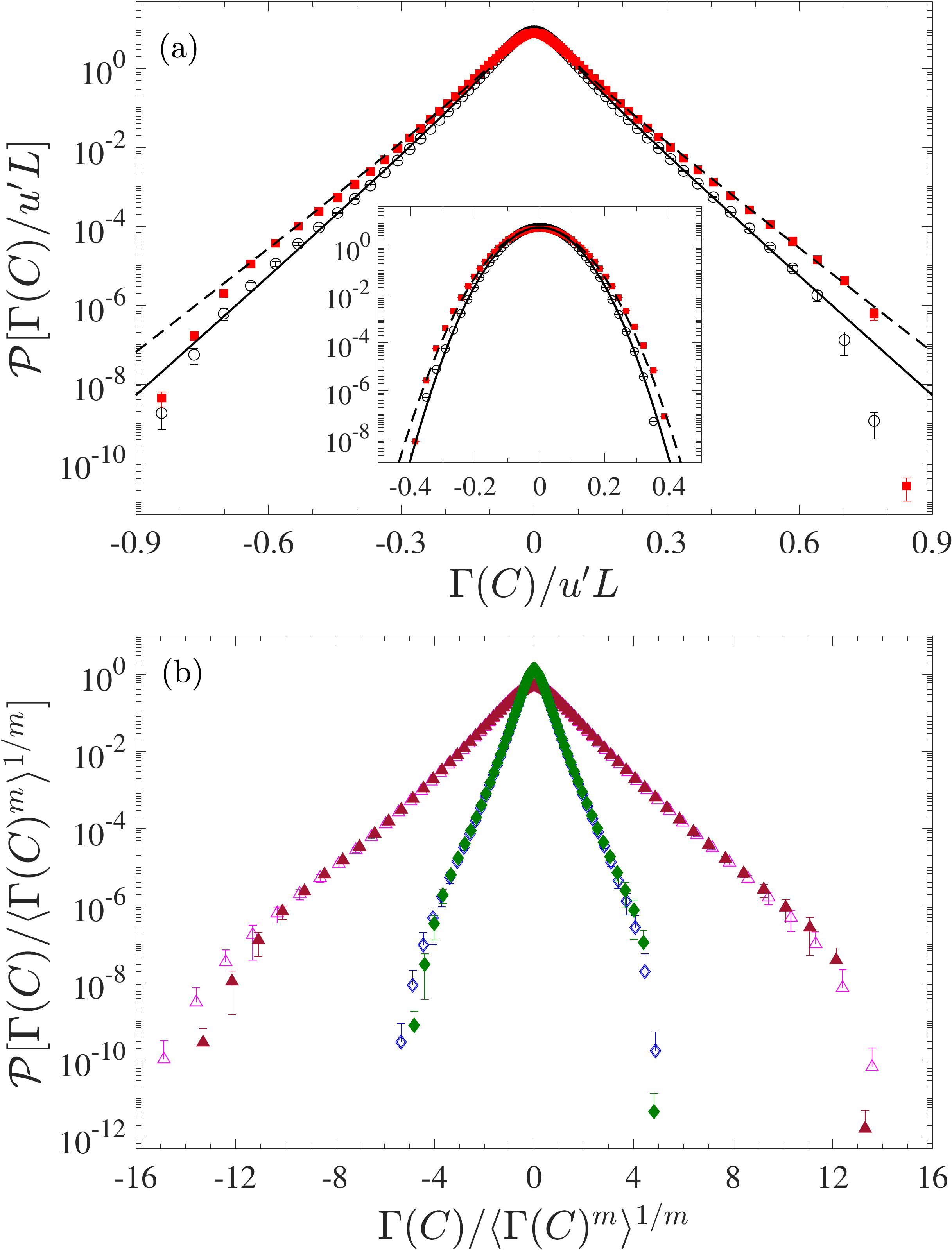}
\protect\caption{
Comparison of circulation PDFs for the non-planar minimal area shown in Fig.~\ref{sg.fig}) and the planar loop with the same area, for $R/\eta = 150$ lying within the inertial range. (a) PDF of $\Gamma(C)/u^\prime L$ for $C_1$ (open circles) and $C_2$ (filled squares). Straight lines are fits $\alpha \exp(-b x)/\sqrt{x}$ in $x \in [0.1, 0.7]$ with $x = |\Gamma(C)|/u^\prime L$, $\alpha = 3.06$, $b = 22.49$ for $C_1$ (solid line) and $\alpha = 2.56$, $b = 19.49$ for $C_2$ (dashed line). The two data points with the lowest probability correspond to $< 2\times 10^{-7} \%$ samples, and are neglected in the fit. The inset shows the corresponding PDF for GRF with Gaussian fit $\alpha_g \exp(-b x^2)$; $\alpha_g = 6.65$, $b_g = 139.4$ for $C_1$ (solid line) and $\alpha = 6.36$, $b = 118.4$ for $C_2$ (dashed line). (b) PDF of standardized circulation $\Gamma(C)/\langle \Gamma(C)^m \rangle^{1/m}$ using $m=2$ (triangles) and $m=8$ (diamonds) for $C_1$ (open symbols) and $C_2$ (filled symbols).  
}
\label{pdfnp.fig}
\end{figure}
   
To build some intuition first, we consider the asymptotic limits of $\Gamma(C_1)$ and $\Gamma(C_2)$. In the dissipative region $R/\eta \approx 1$, $\la \Gamma(C_1)^2 \ra^{1/2} \approx \sqrt{2} \sigma_{\omega} \eta^2 < 1.78 \sigma_{\omega} \eta^2 \approx \la \Gamma(C_2)^2 \ra^{1/2}$, where $ \sigma_{\omega}$ is the standard deviation of the vorticity. In the opposite limit $R/L \approx 1$, the variances approach zero for both $C_1$ and $C_2$. For the intermediate inertial range no such {\it {a priori}} knowledge is available.

Figure \ref{pdfnp.fig}(a) compares the PDF of $\Gamma(C_1)$ and $\Gamma(C_2)$ for a fixed $R$ in the inertial range ($= 150\eta)$. The tails of the PDF (which can be fitted as before) show lower probability for $C_1$ (in 3D) than for $C_2$ (planar); hence all high-order moments of $\gcone$ are smaller than those of $\gctwo$. This rules out  one premise of the Area Rule that the circulation statistics are the same for a given minimal area.

However, the standardized PDFs shown in \ref{pdfnp.fig}(b) for $C_1$ and $C_2$ show good agreement except possibly at the extreme tails which are prone to sampling issues, as noted already. It is noteworthy that the PDF of $\Gamma(C)/\la \Gamma(C)^m \ra^{1/m}$ for both $m=2$ (low order) and $m=8$ (high order) show good collapse at the tails (excepting for the last two data points for reasons noted earlier), which indicates that moments of $\gcone$ and $\gctwo$ match well, when normalized by $\la \Gamma(C)^m \ra^{1/m}$. This says that, circulation equivalence for a given minimal area is restored when the circulation is normalized by internal variables such as its standard deviation or by an appropriate higher order moment.

\begin{figure}
\includegraphics [width=0.5\textwidth,center]{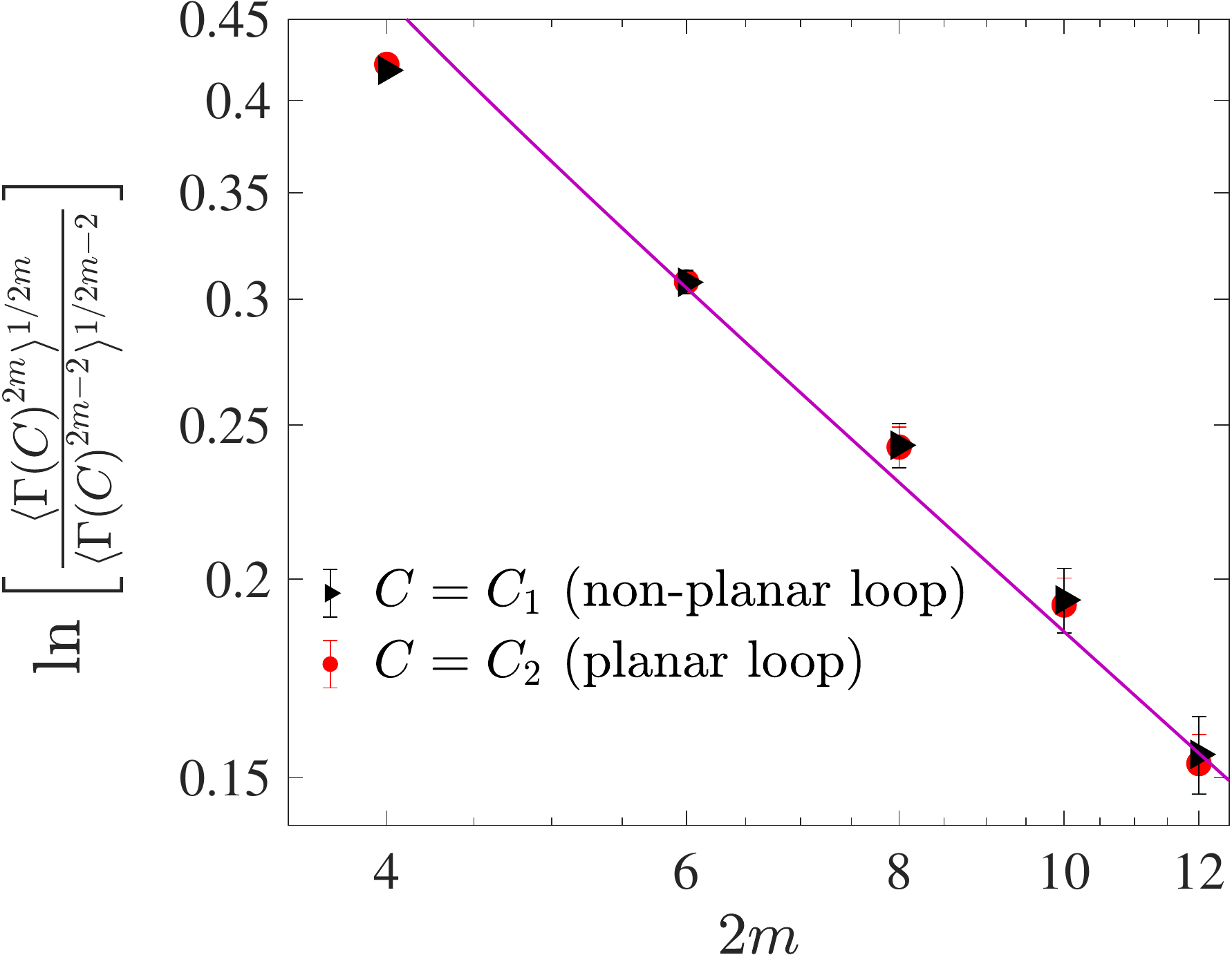}
\protect\caption{
Logarithm of the ratio of circulation moments at successive even orders $2m$ and $2m-2$ plotted against order $2m$ for non-planar loop $C_1$ and planar loop $C_2$, both with same minimal area (see Fig.~\ref{sg.fig}) on log-log scales. Different symbols correspond to DNS data for $C_1$ ($\blacktriangleright$) and $C_2$ ($\colr{\bullet}$) with characteristic loop dimension $R/\eta = 150$ taken in the inertial range (see Fig.~\ref{sg.fig}). 
Solid line is the high-order estimate from Eq.~\ref{stirling.eq} using
$2n=2m-2$, which has the asymptotic slope of $m^{-1}$ coinciding with the GRF slope.
}
\label{momnp.fig}
\end{figure}

In order to examine the order dependence of the normalized PDFs in Fig.~\ref{pdfnp.fig} for loops $C_1$ and $C_2$, we plot the logarithm of the successive even-order moment ratios for an inertial loop $C$ against the moment order in Fig.~\ref{momnp.fig}. The moment ratios for both loops collapse within error bars for orders up to $2m=12$, showing that the ratio of circulation moments are the same for inertial loops $C_1$ and $C_2$. This result is indeed consistent with the collapse of the normalized PDFs in Fig.~\ref{pdfnp.fig}(b). Since the circulation moment ratios determine the shape of the circulation PDF, we infer that its shape is preserved for all circulation amplitudes.
The higher-order moment ratios of this quantity compare well with the universal estimate of Eq.~\ref{stirling.eq} for $2n=2m-2$ shown by the solid line in Fig.~\ref{momnp.fig}, which approaches an $m^{-1}$ scaling for large orders. 


In summary, we have examined the dependence of $\pdf$ on the shape of the loop in both planar and non-planar cases. For the planar case, the tails of the PDFs, corresponding to large circulation amplitudes, can be fitted by $ \alpha \exp(-b |\Gamma(C)|)/\sqrt{|\Gamma(C)|}$ \cite{Migdal20}, which is close to an exponential, unlike for velocity differences where one needs robust stretched exponential fits \cite{krs92}. The higher-order moments themselves do not appear to be loop-independent as posited by Migdal \cite{Migdal19a,Migdal19b,Migdal19c} due to small ($< 5\%$) perimeter corrections; but the normalized higher-order moments indeed are loop-independent and can be approximated by a universal formula, Eq.~\ref{stirling.eq}; the formula is not only loop-independent but also order-dependent. In this sense of normalized moments collapsing, the Area Rule holds well for high-order moments, becoming increasingly independent of the loop geometry. For non-planar loops, the $\Gamma(C)$ PDFs for loops with same minimal area but different geometry differ more substantially, but again the normalized PDFs collapse. This specificity for the non-planar case is possibly due to non-trivial orientation of the local vorticity on the minimal surface \cite{Migdal19a,Migdal19b,Migdal19c}, as we shall explain below. 
 
We have shown that the tails of $\pdf$ for loops with the same minimal area have the same form when the scaling factor is the mean-square circulation, or the locally averaged mean square vorticity or enstrophy density. Since the circulation around loop $C$ of area $A$ and dimension $R$ is related to the locally averaged vorticity over $A$ as $\omega_R = \Gamma(C)/A$, we conclude that in regions with large circulation amplitudes, the ratio of local averages of vorticity $\la \omega_R^m \ra/ \la \omega_R^2 \ra^{m/2}$ should depend only on the minimal area magnitude $A_m = min(A(C))$, which is unique, provided that the contour dimensions fall inside the inertial range. Equivalently the tails of the PDF of the locally averaged enstrophy in large circulation regions could assume a universal form in the inertial range. The implication that the local averages of enstrophy, which are known to be multifractal \cite{krs97}, could assume a universal from in regions with large circulation amplitudes could be important to fully understand the topological roots of vortex dynamics in both classical and quantum turbulence \cite{Moriconi20,Krstulovic,Moriconi21,polanco2021vortex}. Indeed, some aspects of the Area Rule may be more relevant to quantum turbulence, as these references have hinted. In particular, the fact that vorticity is related to the antisymmetric part of the strain rate, or to the transverse velocity increments, which are more intermittent than the symmetric part of the strain rate (or the longitudinal field \cite{RSHT,ISY20}), could have an explanation in the minimal-area scaling of circulation. Elucidation of this possibility is an ongoing effort and will be reported in the future.

\section{Acknowledgments}
We are grateful to A. A. Migdal for continuing and stimulating discussions and P. K. Yeung for his sustained collaboration on the DNS data. We thank A. M. Polyakov particularly for his insight on the Area Rule and G. L. Eyink and E. D. Siggia for uplifting remarks over time. This work is partially supported by the National Science Foundation (NSF), via Grant No. ACI-$1640771$ at the Georgia Institute of Technology. The computations were performed using supercomputing resources provided through the Extreme Science and Engineering Discovery Environment (XSEDE) consortium (which is funded by NSF) at the Texas Advanced Computing Center at the University of Texas (Austin), and the Blue Waters Project at the National Center for Supercomputing Applications at the University of Illinois (Urbana-Champaign).

All relevant data are included in the manuscript.

\end{document}